\documentclass{article}

\PassOptionsToPackage{numbers, compress}{natbib}
\usepackage[preprint]{neurips_2026}

\usepackage[utf8]{inputenc} % allow utf-8 input
\usepackage[T1]{fontenc}    % use 8-bit T1 fonts
\usepackage{hyperref}       % hyperlinks
\usepackage{url}            % simple URL typesetting
\usepackage{booktabs}       % professional-quality tables
\usepackage{amsfonts}       % blackboard math symbols
\usepackage{nicefrac}       % compact symbols for 1/2, etc.
\usepackage{microtype}      % microtypography
\usepackage{xcolor}         % colors

\usepackage{times}
\usepackage{latexsym}
\usepackage{booktabs}
\usepackage{hyperref}

\usepackage{microtype}

\usepackage{inconsolata}

\usepackage{graphicx}
\usepackage{xcolor} % Required for defining colors

\usepackage{amsmath,amssymb,amsfonts}

\usepackage{algorithm}
\usepackage{booktabs}   % For professional table rules (\toprule, \midrule, \bottomrule)
\usepackage{algpseudocode}
\usepackage{xspace}
\usepackage{listings}

\usepackage{amsthm} % Loads the theorem package
\usepackage{graphicx}

\usepackage{enumitem}
\ifx\forsubmission\undefined
\newcommand{\TODO}[1]{\textcolor{red}{TODO: #1}}
\newcommand{\SK}[1]{\textcolor{red}{SK: #1}}

\newcommand{\HK}[1]{\textcolor{blue}{HK: #1}}
\newcommand{\RK}[1]{\textcolor{brown}{RK: #1}}

\newcommand{\fixme}[1]{{\color{red} {#1}}}

\else
\newcommand{\TODO}[1]{\textcolor{red}{}}
\newcommand{\SK}[1]{\textcolor{red}{}}
\newcommand{\FK}[1]{\textcolor{blue}{}}
\newcommand{\HK}[1]{\textcolor{blue}{}}
\newcommand{\RK}[1]{\textcolor{green}{}}

\newcommand{\fixme}[1]{{\color{red} {}}} 

\fi

\newcommand{\squishlist}{
 \begin{list}{$\bullet$}
  { \setlength{\itemsep}{0pt}
     \setlength{\parsep}{3pt}
     \setlength{\topsep}{3pt}
     \setlength{\partopsep}{0pt}
     \setlength{\leftmargin}{1.5em}
     \setlength{\labelwidth}{1em}
     \setlength{\labelsep}{0.5em} } }

\newcommand{\squishlisttwo}{
 \begin{list}{$\bullet$}
  { \setlength{\itemsep}{0pt}
     \setlength{\parsep}{0pt}
    \setlength{\topsep}{0pt}
    \setlength{\partopsep}{0pt}
    \setlength{\leftmargin}{2em}
    \setlength{\labelwidth}{1.5em}
    \setlength{\labelsep}{0.5em} } }

\newcommand{\squishend}{
  \end{list}  }

\newcommand{\betterparagraph}[1]{\noindent \textbf{#1. }}

\newcommand{\method}{\textsc{ETA}\xspace}      % <- rename: deadline-bounded CoT exit
\newcommand{\toolcall}{\texttt{<tool\_call>}}
\newcommand{\thinkend}{\texttt{</think>}}

\usepackage{graphicx}
\graphicspath{{./Figures/}}

\title{ETA: \underline{E}xit on \underline{T}ool \underline{A}rrival in Speculative Agent Loops}
\author{%
\textbf{Saptarshi Mitra}\textsuperscript{1,2},
\textbf{Xuan Zhu}\textsuperscript{1},
 \textbf{Rohith Kukkala\textsuperscript{1}},
 \textbf{Sitao Huang\textsuperscript{2}},
 \textbf{Karthik Vijayan\textsuperscript{1}}
\\
 \textsuperscript{1}Microsoft
 \textsuperscript{2}University of California, Irvine, USA
 \\
 \small{
   \textbf{Correspondence:} \href{mailto:saptarshi.mitra@uci.edu}{saptarshi.mitra@uci.edu}
 }
}

\begin{document}

\maketitle

% %% EDIT YOUR PAPER'S CONTENTS BELOW
\begin{abstract}
Tool-using language agents in explicit thinking modes generate long chains of thought
before acting, making each turn inherently serial: later tokens depend on the tool
observation.
%
% A new paradigm of speculative tool dispatch has emerged that overlaps tool execution with reasoning, but its benefit is limited when reasoning dominates.
Recent speculative tool dispatch methods overlap tool execution with reasoning, but offers limited benefit when reasoning dominates latency.
% For Qwen3-32B on $\tau^2$-bench, tool
% execution accounts for only $4.1\%$ of wall-clock time, capping overlap-only speedup at
% $1.043\times$. In contrast, reasoning generated after the speculative result is ready
% accounts for $38.9\%$ of wall-clock time, corresponding to a $1.64\times$ ceiling.
We present \textbf{ETA},
% (\underline{E}xit on \underline{T}ool \underline{A}rrival), 
% a new paradigm for speculation in agent loops
that treats the speculative fork not as an actuator but as a sensor of the model's decision state, using it to terminate the reasoning of the main decode.
% which repurposes the speculative fork as a sensor of the model's decision state to terminate the reasoning of main decode.
%
It truncates reasoning only when the prediction is stable, agrees with the dispatched call, passes a free probe measuring whether a call will be emitted at all, and the tool result has arrived: a deadline that bounds per-turn regret by the remaining decode cost.
% The method truncates reasoning only when the predicted call is stable,
% matches the dispatched call, passes a one-token free probe for whether the model intends
% to call a tool, and the tool result is available. This deadline bounds per-turn regret
% by the continuation cost.
% The prediction is never inserted into the transcript; the
% model generates its own action, which passes through the original acceptance gate.
% Using one offline-selected rule without retuning, 
\method{} achieves speedups of
$1.33\times$ on $\tau^2$ airline/retail, $1.26\times$ on $\tau^2$ telecom,
and $1.13\times$ on BFCL v4 multi-turn. On $\tau^2$, it decodes $24\%$ fewer
tokens at a nearly unchanged per-token rate without dropping action quality.

\end{abstract}
\section{Introduction}
% \SM{has many repetitions with methods, shorten there or here}
\label{sec:intro}
Tool-using language agents increasingly rely on explicit \emph{thinking} modes.
At each turn, the model generates a chain-of-thought (CoT) and a tool call, then waits for the environment before continuing~\citep{yao2023react}.
Because each observation conditions all subsequent tokens, reasoning and tool execution occur serially.
Their latencies add at every turn and accumulate over the five to fourteen turns required by a typical task.

Speculative tool dispatch reduces this delay~\citep{bai2026spork, huang2025reducing}.
When the main stream emits its first token, the controller forks a request that shares its KV-cache prefix and is constrained to predict a tool call. 
It executes this prediction in parallel with the remaining reasoning and uses the result only if the model eventually emits the same call.
This moves tool execution off the critical path without modifying the model, a draft model, or relaxing the acceptance criterion.
This approach assumes that tool execution is the bottleneck. For reasoning-heavy agents, it often is not.
With Qwen3 in thinking mode on $\tau^2$-bench~\citep{yao2024tau}, tool execution accounts for only $4.1\%$ of wall-clock time, limiting any overlap-only speedup to $1.043\times$.
In contrast, reasoning decoded after the speculative observation is ready accounts for $38.9\%$ of wall-clock time across accepted turns, nine times the budget available from tool overlap.
%
% On the median turn, this redundant decode constitutes $86\%$ of the turn's reasoning.
%
Thus, the greatest opportunity for latency reduction lies in terminating post-observation reasoning, rather than merely overlapping tool execution.

We introduce \method{} to recover this larger budget.
We repurpose the speculative fork from a tool-launching actuator into a sensor of the model’s evolving decision.
Repeated along the CoT, it reveals whether the predicted call has stabilized.
% Dispatch confidence alone is insufficient for this purpose. Because the forced probe
% begins inside a tool call, its confidence is conditional on a call being made and poorly
% predicts whether the turn should call a tool at all (AUC $0.643$ over $7{,}133$ probes).
% We therefore add a one-token \emph{free probe} at the \thinkend{} boundary, which measures
% the marginal probability of \toolcall{} and improves separation to AUC $0.862$.
\method{} truncates reasoning only when the predicted call is stable, matches the tool call already dispatched, passes the free-probe threshold\S\ref{freeprobe}, and the tool result is available.
Finally, tool completion defines the truncation deadline. Stopping earlier only trades decode time for tool wait, while stopping at completion bounds per-turn regret by the remaining decode cost.
%
% The final condition provides a strict deadline: truncating earlier would merely replace decode time with tool wait, whereas truncating after completion bounds per-turn regret by the continuation cost.
%
The predicted call is never inserted into the transcript.
Instead, the model derives its own action from the shortened reasoning, and the original acceptance gate verifies it against the dispatched call.
Thus, the probe determines when to stop reasoning, but the model still produces the final action.

\begin{itemize}[leftmargin=1.2em,itemsep=1pt,topsep=2pt,parsep=0pt]
    \item We identify reasoning that continues after the tool result is ready as the dominant remaining cost.
    % : $f_{\mathrm{tool}}=4.1\%$ (ceiling $1.043\times$) versus $f_{\mathrm{slack}}=38.9\%$ (ceiling $\mathbf{1.64\times}$).

    % \item We show that dispatch confidence cannot safely trigger an exit because it is
    % conditional on a call being emitted. A one-token free probe supplies the missing
    % marginal signal, improving tool/no-tool separation from AUC $0.643$ to $0.862$.

    \item We introduce a deadline-bounded exit rule gated by prediction stability,
    dispatch agreement, and the free probe. Tool readiness prevents additional tool
    wait, no-call continuations are rolled back, and snapshot/restore supports
    state-mutating tools.

    \item 
    % With one offline-selected rule,
    \method{} achieves $\mathbf{1.33\times}$
    speedup on $\tau^2$ airline/retail, $1.26\times$ on $\tau^2$ telecom, and
    $1.13\times$ on BFCL v4 multi-turn~\citep{patil2025bfcl}. On  $\tau^2$,
    it decodes $24\%$ fewer tokens
    % with nearly unchanged wall time per token ($-0.2\%$).
    without a drop in quality.
    Action match changes by $+2.02$\,pp, within the calibrated
    $\delta=3.18$\,pp noise floor.
    % ; we therefore claim non-inferiority, not improvement.
\end{itemize}

\section{Background and Related Works}
\label{sec:background}

\begin{figure}[t]
    \centering
    \includegraphics[width=\linewidth]{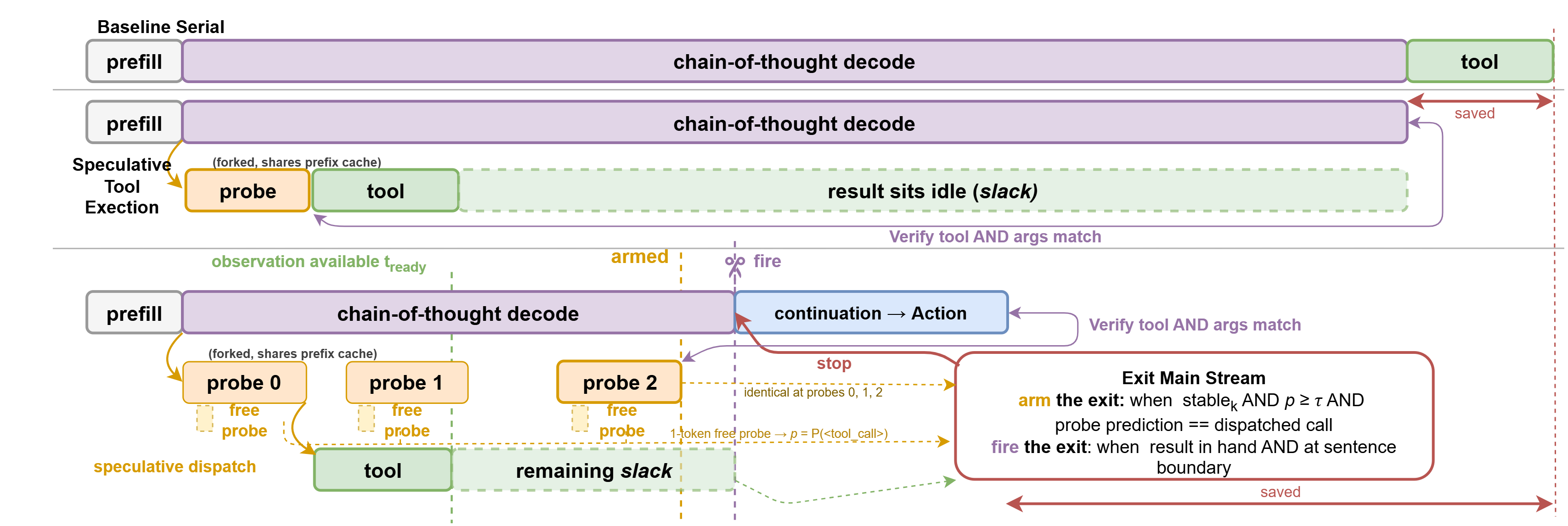}
    \caption{\textbf{\method{} in one turn.} \emph{(a)} The baseline serial execution, speculative tool execution where tool execution is completely overlapped by the CoT and then our mechanism of reusing the speculation to terminate reasoning of main stream early.
    % ~\autoref{fig:overview_mechanism} explains the overall mechanism
    }
    % Measured timeline of a real
    % truncated turn. A probe forked from the main stream predicts the upcoming action
    % and dispatches it early; the observation is ready at $t^{\mathrm{ready}}$, but the
    % main stream would otherwise keep reasoning for another $12$\,s. Once three
    % consecutive probes agree, the free probe reads $p \ge \tau$, and the stable
    % prediction still equals the dispatched call, the exit \emph{arms}; it \emph{fires}
    % at the next sentence boundary, which cannot precede $t^{\mathrm{ready}}$ by
    % construction. \emph{(b)} The decision, with the failure each guard blocks. The
    % probe's prediction is never adopted --- after the cut the model re-derives its own
    % action, which still passes the unmodified acceptance gate, and a turn that declines
    % to act is rolled back onto its original reasoning.}
    \label{fig:overview}
\end{figure}

The tool-wait latency can be eliminated in the agentic ReAct loop by speculatively dispatching the predicted tool call, allowing its execution to overlap with the remaining reasoning.
Now we ask \emph{what this overlap leaves behind?}
The answer motivates the rest of the section: once tool-wait latency is hidden, the dominant remaining cost is the reasoning that continues even after the tool-result is already available.
Our contribution is to reuse the speculation as a \emph{readout of the model's decision state}. 
We use this readout to terminate reasoning early, subject to a deadline that makes the operation latency-safe by construction.
The overall mechanism is illustrated in ~\autoref{fig:overview}.

% We begin from the standard remedy for tool-wait in agentic decoding --- speculatively
% dispatching a predicted tool call so that its execution overlaps the remaining
% reasoning --- and then ask what that overlap leaves behind. The answer motivates the
% rest of the section: on reasoning-heavy agent workloads the overlap saturates almost
% immediately, and the dominant remaining cost is not the tool at all but reasoning that
% continues after the tool result is already in hand. Our contribution is to reuse the
% speculative machinery as a \emph{readout of the model's decision state}, and to use
% that readout, under a deadline that makes the operation latency-safe by construction,
% to stop the reasoning early.

% \insertWideFigureShrink{fig_overview}

% ===========================================================================
% \subsection{Problem setup}
% \label{sec:method:setup}

\paragraph{The ReAct agentic loop.}
A Large Reasoning Model (LRM) agent operates under policy $\pi$.
An episode comprises $H$ turns, indexed by $t = 1,\dots,H$.
At turn $t$ the policy is conditioned on a context $C_t$ (system prompt, user request, and all previous
tool observations) and greedily decodes:
\begin{equation}
    \underbrace{r_t = (r_{t,1},\dots,r_{t,L_t})}_{\text{chain of thought}}
    \;\oplus\;
    \thinkend
    \;\oplus\;
    \underbrace{a_t}_{\text{action}},
    \qquad
    a_t \in \mathcal{A} \cup \{\varnothing\},
    \label{eq:react_loop}
\end{equation}
where an action $a_t = (u_t, v_t)$ consists of a tool name $u_t$ with its arguments $v_t$, and $a_t = \varnothing$ denotes a turn that terminates in a natural-language response to the user rather than a tool-call.
A tool-call takes $T^{\mathrm{tool}}_t$ seconds, and appends the resulting observation $o_t$ to $C_{t+1}$.
% When $a_t \neq \varnothing$ the environment
% executes it, taking $T^{\mathrm{env}}_t$ seconds and returning an observation $o_t$
% that is appended to $C_{t+1}$.
A serial ReAct turn therefore costs: 
\begin{equation}
    T^{\mathrm{ser}}_t
    = T^{\mathrm{pre}}_t
    + \underbrace{(L_t + m_t)\,\bar{d}}_{\text{decode: reasoning + action}}
    + \underbrace{T^{\mathrm{tool}}_t}_{\text{tool wait}},
    \label{eq:serial}
\end{equation}
where $T^{\mathrm{pre}}_t$ denotes prefill time, $\bar{d}$ is the mean per-token decode latency and $m_t$ is the tool-call length in tokens.
Here $o_t$ conditions every subsequent token, so $T^{\mathrm{tool}}_t$ can not start until decoding completes.
\betterparagraph{Self-speculative tool dispatch}Recent works SPORK, SPAgent~\citep{bai2026spork,huang2025reducing} remove tool wait from the critical path by predicting and dispatching tool calls before the model emits them with reasoning.
% We adopt the self-speculative dispatch mechanisms from previous works ~\citep{bai2026spork}~\citep{huang2025reducing}.
%
Upon the main stream's first token, the controller forks a second request that shares the KV-cache prefix (prefill is a cache-hit) and is forced to produce a tool call, yielding a predicted action $\hat a_t$.
If the fork's logprob-based confidence exceeds a threshold $\theta$, $\hat a_t$ is dispatched immediately, in parallel with the main stream’s remaining decode.
Once the main stream eventually emits its own action $a_t$, the precomputed observation is used only if $\hat a_t \equiv a_t$ under exact tool-name and argument matching; otherwise, it is discarded and the tool re-executed serially.
Speculation is therefore lossless relative to serial execution because the agent never observes a mispredicted call.
%
% Let $\alpha$ denote the fraction of tool-emitting turns for which speculation is accepted.

\betterparagraph{Other related works} More broadly, tool-speculation methods differ in how they predict future invocations and how far ahead they speculate.
Speculative Actions~\citep{ye2025speculative} uses a smaller draft model to predict the next action from an observation and execute it early.
% , whereas SpecHop~\citep{saberi2026spechop} maintains asynchronously verified branches across multiple retrieval hops.
%
PASTE~\citep{sui2026paste} learns recurring tool-use patterns from historical traces and replays them through a middleware interceptor.
\citep{nichols2025spectool} instead draft tool calls within the serving engine using a smaller model and keep candidate sequences resident to avoid repeated prefill. 
IdleSpec~\citep{choi2026idlespec} uses the same idle periods to generate speculative \emph{plans}, trading additional computation for accuracy rather than latency.
MORI~\citep{xia2026idleness} addresses a complementary systems cost by offloading the KV cache during tool waits rather than shortening the critical path.
A separate line of work reduces chain-of-thought length~\citep{sui2025stopoverthinking}. 
Approaches include fixed-token budget forcing~\citep{muennighoff2025s1}, reinforcement learning for length pruning~\citep{hou2025thinkprune}, and dynamic exit rules.
DEER generates trial answers at reasoning transitions and exits when confidence is high~\citep{yang2026deer}.
Dynasor/Certaindex probes the model at fixed token intervals, extracting its current answer without altering the main stream, and uses the resulting certainty signal to allocate test-time compute~\citep{fu2024certaindex}.

\section{Methodology}
\label{sec:method}
When the tool execution lies on the serial critical path, its runtime fraction is $f_{\mathrm{tool}} = \frac{1}{T}\sum_t T^{\mathrm{tool}}_t$.
Even perfect tool overlap removes only $f_{\mathrm{tool}}$ of runtime, limiting speedup to $1/(1 - f_{\mathrm{tool}})$.
On our primary workload ($\tau^2$-bench, Qwen3-32B in thinking mode, and a $2.0$\,s tool-call floor) with RTX 6000 Pro GPU, we measure $f_{\mathrm{tool}} = 4.1\%$, implying a speedup ceiling of $\mathbf{1.043\times}$.
For reasoning models that decode hundreds of tokens per turn, tool execution is therefore not the real bottleneck.
This section examines what remains on the critical path after tool wait is hidden.

\begin{figure}[t]
    \centering
    \includegraphics[width=\linewidth]{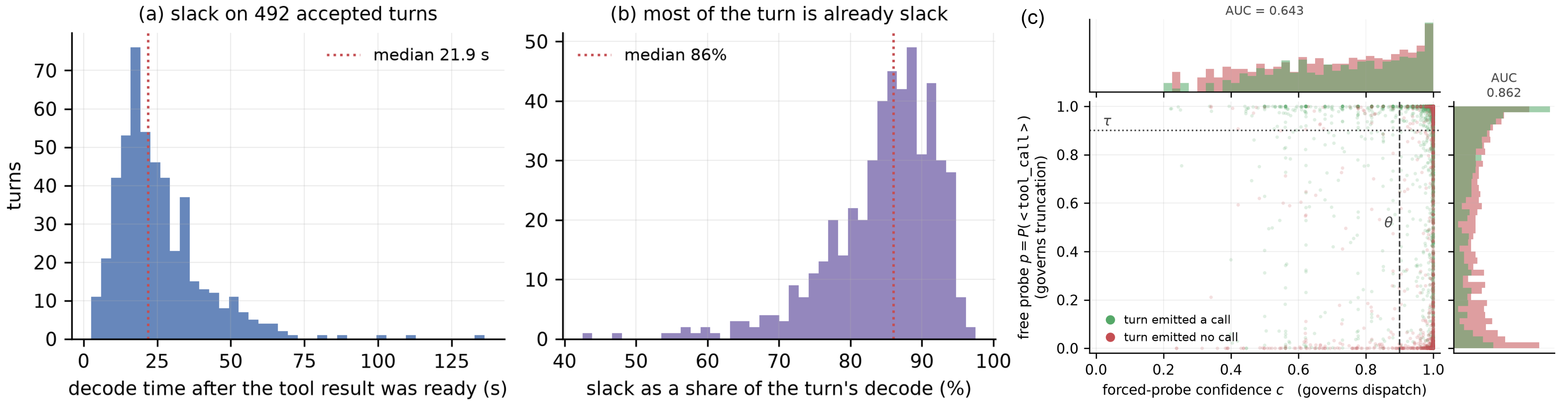}
    \caption{
    \emph{(a)} Per-turn slack $\sigma_t$ across accepted speculative turns, measuring decode time after the observation was available. \emph{(b)} Slack normalized by turn-level decode time, with a median of 86\%. \emph{(c)}
    Forced-probe conf. $c$ is nearly blind to  tool/no-tool decision:
    % Forced-probe confidence $c$ fails to distinguish tool-call from no-tool turns: 
    dispatch threshold $\theta$ admits $93.4\%$ of tool turns but also $87.4\%$ of no-tool turns. The truncation threshold $\tau$ admits $82.3\%$ against $28.5\%$. }
    \label{fig:slack}
\end{figure}

\betterparagraph{Exploiting the post-observation reasoning slack}Now for each overlapped turn, the slack is $\sigma_t = \bigl(e_t - t^{\mathrm{ready}}_t\bigr)$, where $e_t$ is when the main stream finishes decoding turn $t$, and $t^{\mathrm{ready}}_t$ is when the observation for $a_t$ becomes available in the speculative fork.
The aggregate slack fraction $f_{\mathrm{slack}} = \frac{1}{T}\sum_t \sigma_t$ is the optimization budget targeted in this work.
In the same setup, $f_{\mathrm{slack}} = 38.9\%$, implying a speedup ceiling of $S = \frac{1}{1 - f_{\mathrm{slack}}} = \mathbf{1.64\times}$, with a $9\times$ larger optimization budget than tool-execution overlap.
Per accepted turn, the median slack is $21.9$\,s and, as a share of that turn's decode, the median is $86\%$ (~\autoref{fig:slack}a-b).
Thus the observation of a turn is typically ready while most of the reasoning remains to be decoded, which we plan to exploit.

Detecting when the model has \emph{committed} to an action lets us terminate the remaining reasoning.
The speculative fork provides this signal: although designed to predict \emph{which tool to launch}, probing it throughout the CoT reveals \emph{whether the prediction has stabilized}. 
We thus repurpose the fork from an actuator into a sensor.
Truncation is permitted only once the observation from the tool becomes available.
Stopping earlier cannot reduce latency because it merely replaces decode time with an equal tool-wait, while risking an incorrect decision.

% \insertWideFigureShrink{fig7_slack}{}

\subsection{\method{} Mechanism}
\label{sec:method:mechanism}

As \autoref{eq:react_loop} shows, not every turn ends with a tool call.
We must therefore distinguish assertions from tool calls before arming the exit mechanism.

\betterparagraph{Free Probe}
\label{freeprobe}
Prior work uses confidence over the probe’s tool-name span to decide whether to speculatively dispatch, but this confidence cannot serve as an exit signal.
Because the forced-probe prompt ends \emph{inside} a tool call (below, left), its continuation is restricted to tool names.
Thus, $c^{(j)} \approx P_\pi(u = \hat u^{(j)} \mid \text{a call is emitted})$ measures confidence in the predicted tool \emph{conditional} on a call, but provides no evidence that a call should occur.
When the correct outcome is an assertion, the probe cannot be correct, yet may remain highly confident in the tool name it is forced to generate.

% The dispatch confidence cannot be reused as an exit signal, for two independent
% reasons. First, the forced probe's prompt terminates \emph{inside} a tool call
% (below, left), so the support of its continuation is the set of tool names and
% $c^{(j)} \approx P_\pi(u = \hat u^{(j)} \mid \text{a call is emitted})$: a
% \emph{conditional} quantity, structurally silent about the event it conditions on. On a
% turn whose correct outcome is a plain answer the probe is not merely wrong but
% \emph{incapable of being right}, and still reports high confidence in the name it was
% forced to invent. Second, $c$ is a minimum over a fixed window of the name span. On one
% turn the probe hallucinated the order id \texttt{\#W5362222} in place of
% \texttt{\#W5362037} at $c = 0.99998$, because the $16$-token window is consumed by the
% tool name, the JSON boilerplate and the shared prefix \texttt{\#W536}, placing the
% divergence outside it --- \emph{confidence over a name span cannot supervise an
% argument}. The two guards we adopt instead follow directly: exact-argument stability
% (\S\ref{sec:method:rule}), and a marginal tool-call probability.

\medskip
\noindent
\begin{minipage}[t]{0.48\linewidth}
\centering{\footnotesize\textbf{forced probe}: \emph{which} tool?}
\vspace{2pt}\hrule\vspace{3pt}
{\scriptsize\ttfamily
\ldots<|im\_start|>assistant\\
<think>\\
Okay, let's see. The user is James\ldots\\
</think>\\[2pt]
<tool\_call>\\
\{"name": "\hfill $\leftarrow$ ends here
}
\vspace{3pt}\hrule
\end{minipage}\hfill
\begin{minipage}[t]{0.48\linewidth}
\centering{\footnotesize\textbf{free probe}: \emph{whether} a tool?}
\vspace{2pt}\hrule\vspace{3pt}
{\scriptsize\ttfamily
\ldots<|im\_start|>assistant\\
<think>\\
Okay, let's see. The user is James\ldots\\
</think>\\[2pt]
\phantom{<tool\_call>}\\
\hfill $\leftarrow$ ends here
}
\vspace{3pt}\hrule
\end{minipage}
\medskip

High probe confidence is insufficient for safely terminating the reasoning.
% Terminating the main stream based on high probe confidence can yield an incorrect action.
% If we terminate the main stream based on high confidence, it will be wrong.
For solution, we create an additional probe by removing the final two prompt lines (right).
This places the model at a genuine decision point: its next token is either \toolcall{} or the first token of the assertion.
We request \emph{exactly one token} with \texttt{logprobs=20} and measure the probability mass assigned to \toolcall{}.
% Deleting the last two lines puts the model at a genuine fork: its next token is either
% the special token \toolcall{} or the first token of prose. We request \emph{exactly one
% token} with \texttt{logprobs=20} and read off the mass on \toolcall{} (id
% \texttt{151657} for Qwen3):
\begin{equation}
    p^{(j)} \;=\; \exp\ell\bigl(\toolcall\bigr)\cdot
    \mathbb{1}\bigl[\toolcall \in \text{top-}20\bigr].
    \label{eq:ptool}
\end{equation}

This probe costs a single decode step for prefix cache hit.
Although both probes read the same model state, $c$ identifies \emph{which tool} from inside the call (to dispatch the tool), whereas $p$ determines \emph{whether to call a tool} at the \thinkend{} boundary.
%
% That is the entire computation, and because the prompt is a prefix of what the server
% has just decoded it is a cache hit costing one decode step. The two probes therefore
% read the same state at different positions and answer different questions --- $c$ from
% inside the call, $p$ at the \thinkend{} boundary --- and fail on different turns.
%
In ~\ref{appendix:example}, Turn 4 of a worked example demonstrate the utility of devising the free probe, which prevents an incorrect action due to early exit.
~\autoref{fig:slack}(c) confirms the same effect at scale.
Across all $7133$ probes of the $\tau^2$-suite, $c$ poorly distinguishes turns that end in an action from those that do not (AUC $0.643$), whereas $p$ separates them effectively (AUC $0.862$).

% Figure~\ref{fig:signals} shows the same effect at scale: across all $7{,}133$ probes of
% the suite, $c$ is near-identical on turns that do and do not end in an action
% (median $1.0000$ vs $0.9997$; AUC $0.643$), whereas $p$ separates them
% (median $0.9996$ vs $0.1936$; AUC $0.862$).

% \begin{figure}[t]
%     \centering
%     \includegraphics[width=0.7\linewidth]{figs/fig9_signals}
%     \caption{\textbf{The dispatch confidence is nearly blind to the question the
%     truncation decision asks.} All $7{,}133$ probes of the instrumented control arm,
%     plotted as forced-probe confidence $c$ against free-probe $p = P(\toolcall)$ and
%     coloured by whether the turn ultimately emitted a tool call, with marginals. Both
%     populations pile onto $c \approx 1$ --- the forced probe's prompt mandates a call,
%     so its confidence is conditional on an event it cannot question --- and the
%     dispatch threshold $\theta$ admits $93.4\%$ of tool turns but also $87.4\%$ of
%     no-tool turns. The truncation threshold $\tau$ admits $82.3\%$ against $28.5\%$.}
%     \label{fig:signals}
% \end{figure}

\betterparagraph{The deadline-bounded exit mechanism}
Probes are forked at CoT positions $n_0 = 0 < n_1 < \cdots < n_J$, spaced $\Delta = 50$ tokens apart and snapped to sentence boundary with max $J = 15$ probes per turn (~\ref{app:cadence}).
Let $\hat a^{(j)} = (\hat u^{(j)}, \hat v^{(j)})$ denote the action predicted by probe $j$.
We write $\hat a^{(j)}\equiv a^{(k)}$ when their tool names and arguments match exactly, following the acceptance-gate criterion, and use $\check a_t$ for the dispatched call (whenever the dispatch threshold was satisfied).
We define stability and the \textcolor[HTML]{D79B00}{arm}ing condition as follows:
\begin{align}
    \mathrm{stable}_k(j)
        &= \mathbb{1}\Bigl[\hat a^{(j)} \equiv \hat a^{(j-1)} \equiv \cdots
                            \equiv \hat a^{(j-k+1)}\Bigr],
    \label{eq:stable}\\[2pt]
    A_j
        &= \mathrm{stable}_k(j)
        \;\wedge\;
        \bigl(p^{(j)} \ge \tau\bigr)
        \;\wedge\;
        \bigl(\hat a^{(j)} \equiv \check a_t\bigr).
    \label{eq:arm}
\end{align}
Arming is necessary but insufficient.
Let $\mathrm{ready}_t(n)$ indicate that speculative execution has completed by the time token $n$ is produced, and let $\mathrm{bnd}(\cdot)$ indicate a sentence boundary.
We \textcolor[HTML]{9673A6}{fire} the cut point at:
\begin{equation}
    n^{\dagger}
    = \min\Bigl\{\, n \;:\;
        \mathrm{ready}_t(n) \;\wedge\; \mathrm{bnd}(r_{t,1:n})
        \;\wedge\; \bigl(\exists\, j \le J:\; A_j \wedge n \ge n_j\bigr) \Bigr\}.
    \label{eq:fire}
\end{equation}
At $n^{\dagger}$, we cooperatively stop the main stream and conclude the turn using:
\begin{equation}
    a_t \sim \pi\bigl(\cdot \,\big|\, C_t,\; r_{t,1:n^{\dagger}} \oplus \thinkend\bigr).
    \label{eq:continuation}
\end{equation}
We append only \thinkend{} and never inject $\check a_t$ into the main stream.
Thus, truncation cannot force the model to emit an unintended action.
\autoref{fig:overview_mechanism} illustrates the mechanism, while Alg. \autoref{alg:turn} specifies the complete procedure.
% ~\autoref{fig:overview_mechanism} and Alg. \autoref{alg:turn} explains the overall mechanism in detail.
%
~\ref{app:exit_points} provides additional rationale for the exit mechanism. 

\section{Evaluation and Results}
\label{sec:results}

% \paragraph{Setup.}
Our setup uses Qwen3-32B in thinking mode, served by vLLM\cite{kwon2023efficient}, with a $2.0$\,s tool-call latency floor. We compare five methods that differ only in the mechanism under test: \textbf{serial} (no speculation), \textbf{SPORK (minimal)}~\citep{bai2026spork}, \textbf{SPORK} (confidence-gated retry probing; strongest baseline), \textbf{ours (off)} (full instrumentation: pays the extra cost but takes no benefit), and \textbf{ours (on)} (truncation enabled).

% Qwen3-32B with thinking enabled, served by vLLM, a $2.0$\,s latency floor on every tool call.
% % , and one task per process invocation against a freshly reset environment (App.~\ref{app:protocol} explains why batching tasks together is not admissible here).
% Five methods, identical in every respect except the mechanism under test: \textbf{serial} (no speculation); \textbf{SPORK}(minimal) and \textbf{SPORK}\cite{bai2026spork} (with confidence-gated retry probing); \textbf{ours (off)}, which runs our full instrumentation but early-exit disabled;
% % --- dense probing, free-probe logprobs, the exit rule evaluated at every probe --- with $\tau = 1.1$, which no probability can satisfy, so the exit arms zero times on an otherwise identical code path;
% and \textbf{ours (on)}, the same arm with truncation enabled.
% %

% The \textbf{ours (off)} control is what makes the attribution clean: it pays every cost
% the mechanism introduces ($8.9$ probes/turn against D1+D2's $1.3$) and takes none of the
% benefit, so \textbf{ours (off) $\to$ ours (on)} isolates the truncation action alone.
% All run and analysis commands are listed in App.~\ref{app:repro}.

% ===========================================================================
% \subsection{Main result}
% \label{sec:results:main}

Table~\ref{tab:methods} compares the five methods across all 155  $\tau^2$-bench airline and retail tasks, comprising 692 gold actions.
Table~\ref{tab:chain} reports every pairwise speed and quality contrast, with the three \emph{zero-by-construction} comparisons first.
The treatment effect on action match is smaller than the largest calibration contrast expected to be zero  ($+2.02$ vs. $\delta = 3.18$\,pp), while every lower bound remains above the non-inferiority margin $-\delta$ (~\ref{app:noninf}).

% The treatment moves action match less than the largest should-be-zero calibrator does ($+2.02$ against $\delta = 3.18$\,pp), and every  lower bound clears $-\delta$ (Appendix~\ref{app:noninf}).

\begin{table}[h]
\centering\footnotesize
\setlength{\tabcolsep}{5pt}
\caption{Five methods, 155 $\tau^2$ tasks, 692 golden actions.}
\label{tab:methods}
\begin{tabular}{@{}lrrrrrrr@{}}
\toprule
method & wall (s) & P50 & P95 & action match & turns & decode tok & ms/tok \\
\midrule
serial          & 31\,505 & 163 & 510 & 337/692 (48.7\%) & 785 & 598\,968 & 52.60 \\
SPORK              & 32\,495 & 174 & 478 & 359/692 (51.9\%) & 824 & 635\,477 & 51.13 \\
SPORK (retry)           & 31\,675 & 170 & 457 & 351/692 (50.7\%) & 839 & 620\,419 & 51.05 \\
ours (off)      & 33\,421 & 171 & 495 & 334/692 (48.3\%) & 797 & 644\,407 & 51.86 \\
\textbf{ours (on)} & \textbf{23\,743} & \textbf{125} & \textbf{412}
                & \textbf{365/692 (52.7\%)} & 871 & \textbf{452\,521} & 52.47 \\
\bottomrule
\end{tabular}
\end{table}

\begin{table}[h]
\centering\footnotesize
\setlength{\tabcolsep}{4pt}
\caption{
Every contrast on the same 155 tasks.
% Paired task-level bootstrap, 20\,000 resamples.
Quality is action match
% the last column is the one-sided $95\%$
% lower bound used for the non-inferiority test.
}
\label{tab:chain}
\begin{tabular}{@{}lrlr@{\hskip 10pt}rlr@{}}
\toprule
& \multicolumn{3}{c}{wall-clock} & \multicolumn{3}{c}{action match} \\
\cmidrule(r){2-4}\cmidrule(l){5-7}
contrast & saved & 95\% CI & speedup & $\Delta$ pp & 95\% CI & LCB \\
\midrule
\multicolumn{7}{@{}l}{\emph{zero by construction : these calibrate the apparatus}}\\
serial $\to$ SPORK        & $-3.1\%$ & [$-14.7$, $+6.9$] & 0.970$\times$
                       & $+3.18$ & [$-1.22$, $+7.55$] & --- \\
SPORK $\to$ SPORK(retry)         & $+2.5\%$ & [$-5.0$, $+9.7$]  & 1.026$\times$
                       & $-1.16$ & [$-5.02$, $+2.68$] & --- \\
SPORK(retry) $\to$ ours (off) & $-5.5\%$ & [$-16.1$, $+4.1$] & 0.948$\times$
                       & $-2.46$ & [$-7.79$, $+3.05$] & --- \\
\midrule
\multicolumn{7}{@{}l}{\emph{treatment}}\\
\textbf{SPORK(retry) $\to$ ours (on)} & $\mathbf{+25.0\%}$ & \textbf{[$+16.1$, $+32.8$]}
                       & $\mathbf{1.334\times}$ & $\mathbf{+2.02}$ & --- & $-2.22$ \\
ours (off) $\to$ ours (on) & $+29.0\%$ & [$+21.6$, $+35.5$] & 1.408$\times$
                       & $+4.48$ & --- & $+0.83$ \\
\textbf{serial $\to$ ours (on)} & $\mathbf{+24.6\%}$ & [$+15.4$, $+32.6$] & $\mathbf{1.327\times}$
                       & $\mathbf{+4.05}$ & --- & $+0.78$ \\
SPORK $\to$ ours (on)     & $+26.9\%$ & [$+18.4$, $+34.3$] & 1.369$\times$
                       & $+0.87$ & --- & $-3.11$ \\
\bottomrule
\end{tabular}
\end{table}

% Three readings, in the order a sceptical reader should take them.

% \textbf{(i) The speedup is real and is attributable to truncation.} Every calibrator
% interval contains zero; every treatment interval excludes it. The headline is
% $\mathbf{1.334\times}$ over production D1+D2 --- the conservative choice, since it
% charges the mechanism for its own probing overhead --- and the cleanest causal statement
% is $\mathbf{1.408\times}$ from ours\,(off), which holds the probe cadence, the logprob
% requests and the code path fixed and changes only whether the exit may fire.

% \textbf{(ii) Quality does not move by more than the apparatus can resolve.} The largest
% magnitude among the should-be-zero comparisons is $\delta = 3.18$\,pp, so differences
% below that are not resolvable by this measurement. Every treatment lower bound clears
% $-\delta$; two clear zero. Crucially, \emph{turning truncation on perturbs action match
% less than adding D1 to a serial baseline does}. We claim non-inferiority at
% $\delta = 3.18$\,pp, not improvement, notwithstanding the positive point estimates
% (App.~\ref{app:noninf} gives the bootstrap design and the replicate schedule that would
% tighten $\delta$).

% \textbf{(iii) The win is fewer tokens, not faster tokens.} 
Relative to serial, wall time and decoded tokens decrease by $24.6\%$ and  $24.4\%$, respectively, while wall time per token changes by only $-0.2\%$ (Table~\ref{tab:methods}, last column).
% ;Figure~\ref{fig:tokens}).
%
This confirms that the speedup comes from decoding fewer tokens rather than better tool overlap. In contrast, SPORK reduces wall time per token by $2.8\%$ and $2.9\%$, but emits $6.1\%$ and $3.6\%$ more tokens, negating their gain.
% Wall time falls $24.6\%$ against serial while decoded tokens fall $24.4\%$ at a wall-per-token that moves $-0.2\%$ (Table~\ref{tab:methods}, last column; Figure~\ref{fig:tokens}).
% %
% This is the load-bearing check: an arm that merely overlapped tool time better would move ms/token, and SPORK configs do exactly that ($-2.8\%$, $-2.9\%$); and are then outrun by emitting $6.1\%$ and $3.6\%$ more tokens.
%
% Our method hides \emph{more} tool time than any other ($1\,132.6$\,s over $567/716$ accepted turns) and still shows $-0.2\%$, because continuations and rollbacks are wall-clock that produces no tokens.
%
% Since the token count is not invariant, the conventional overlap identity does not apply to our arm and we do not quote it (App.~\ref{app:accounting}).
% \begin{figure}[t]
%     \centering
%     \includegraphics[width=\linewidth]{figs/fig1_tokens_vs_wall}
%     \caption{\textbf{Every arm lies on one decode-rate line; only the position along it
%     changes.} \emph{(a)} Total wall-clock against total decoded tokens, one point per
%     arm, with the serial decode rate ($52.60$\,ms/token) as reference. \emph{(b)} The
%     same data as wall per decoded token relative to serial. Overlap moves an arm
%     \emph{off} the line (D1, D1+D2 at $-2.8\%$, $-2.9\%$) but then moves it \emph{along}
%     the line in the wrong direction by emitting more tokens; our arm moves along the line
%     in the right direction. The mechanism is a work reduction, not a rate improvement.}
%     \label{fig:tokens}
% \end{figure}
Our speedup is not carried by outliers: $118/155$ tasks $(76\%)$ are faster with a median of $1.28\times$, and P90 latency falls $393 \to 267$\,s $(-32\%)$ (~\ref{app:mech}).

% ===========================================================================
% \subsection{The speedup is distributed, not carried by outliers}
% \label{sec:results:dist}

% Per-task speedup against D1+D2: P05 $0.43\times$, P25 $1.01\times$, median
% $\mathbf{1.28\times}$, P75 $1.91\times$, P95 $4.40\times$. $118/155$ tasks $(76\%)$ are
% faster and $34$ are at least $2\times$ faster. The whole latency distribution shifts,
% including the tail: P90 $393 \to 267$\,s $(-32\%)$, P95 $457 \to 412$\,s.

% The exit fires on $589$ of $871$ turns $(67.6\%)$, cutting chain-of-thought from a
% median of $545$ tokens (D1+D2) to $254$ on the turns it truncates. It behaves as an
% accuracy selector rather than a blunt shortener: speculations on truncated turns are
% accepted $93.2\%$ of the time $(549/589)$ against $7.6\%$ $(18/238)$ on turns it
% declines to truncate, which is why overall acceptance \emph{rises} to $68.6\%$
% $(567/827)$ from D1+D2's $65.2\%$ despite reasoning for $27\%$ fewer tokens
% (App.~\ref{app:mech}).

% ===========================================================================
\betterparagraph{Results on other tasks}
The operating point ($k = 3$, $\tau = 0.9$, dispatch-match) was selected offline once and frozen before any full task run; no workload in Table~\ref{tab:general} was
retuned (App.~\ref{app:oracle}).
% \subsection{Generalisation, and a boundary}
% \label{sec:results:general}
% The operating point ($k = 3$, $\tau = 0.9$, dispatch-match) was selected offline once
% and frozen before any suite-scale run; no workload in Table~\ref{tab:general} was
% retuned (App.~\ref{app:oracle}).
\begin{table}[h]
\centering\footnotesize
\setlength{\tabcolsep}{4pt}
\caption{Three workloads, one frozen rule; paired task-level bootstrap.
}
% \caption{Four workloads, one frozen rule; paired task-level bootstrap. Quality is
% against D1+D2 and its metric differs by workload: action match,
% \texttt{ENV\_ASSERTION}, turns passed, EM. $^{\dagger}$Table~\ref{tab:chain}'s intervals
% as ratios. $^{\ddagger}$$n = 30$ for this contrast. $^{\S}$no D1+D2 arm on this subset.}
\label{tab:general}
\begin{tabular}{@{}llll l@{}}
\toprule
& & \multicolumn{2}{c}{speedup [95\% CI]} & \\
\cmidrule(lr){3-4}
workload & $n$ & vs.\ baseline & vs.\ serial & quality \\
\midrule
$\tau^2$ airline+retail & 155
  & $\mathbf{1.334\times}$ [1.19, 1.49]
  % $^{\dagger}$
  & $1.327\times$ [1.18, 1.48]
  % $^{\dagger}$
  & $\mathbf{+2.02}$ pp, non-inf. \\
$\tau^2$ telecom-solo & 40
  & $1.258\times$ [1.143, 1.391]
  & $\mathbf{1.326\times}$ [1.194, 1.474] & $\mathbf{+8.70}$ pp, non-inf. \\
BFCL v4 multi-turn & 200
  & $\mathbf{1.134\times}$ [1.071, 1.215]
  & $1.133\times$ [1.022, 1.258]
  % $^{\ddagger}$
  & $\mathbf{+2.59}$ pp, non-inf. \\
% HotpotQA & 65
%   & ---$^{\S}$
%   & $1.063\times$ [0.94, 1.21] & EM $-1$, [$-8$, $+6$] \\
\bottomrule
\end{tabular}
\end{table}
The method generalizes to $\tau^2$ telecom-solo, a diagnose-act-verify workload averaging $13.8$ turns per task, reaches $1.258\times$ against baseline.
On stateful BFCL tasks, it achieves a $1.134\times$ speedup, with $11.8\%$ lower wall time and $14.3\%$ fewer tokens.
Quality is non-inferior or statistically unchanged throughout, and BFCL wrong-state failures decrease from $47 \to 37$.
\section{Conclusions}
\label{sec:conclusion}
After speculative dispatch hides tool latency, post-observation reasoning becomes the dominant remaining cost.
ETA uses the speculative fork to detect action commitment and safely truncate this reasoning under a bounded-latency deadline.
ETA preserves non-inferior action quality while achieving $1.33\times$, $1.26\times$, and $1.13\times$ speedups across three workloads against state-of-the-art baseline.

\bibliographystyle{plainnat}
\bibliography{ref} % Points to your ref.bib file

% \section*{References}

% References follow the acknowledgments in the camera-ready paper. Use unnumbered first-level heading for
% the references. Any choice of citation style is acceptable as long as you are
% consistent. It is permissible to reduce the font size to \verb+small+ (9 point)
% when listing the references.
% Note that the Reference section does not count towards the page limit.
% \medskip

% {
% \small

% [1] Alexander, J.A.\ \& Mozer, M.C.\ (1995) Template-based algorithms for
% connectionist rule extraction. In G.\ Tesauro, D.S.\ Touretzky and T.K.\ Leen
% (eds.), {\it Advances in Neural Information Processing Systems 7},
% pp.\ 609--616. Cambridge, MA: MIT Press.

% [2] Bower, J.M.\ \& Beeman, D.\ (1995) {\it The Book of GENESIS: Exploring
%   Realistic Neural Models with the GEneral NEural SImulation System.}  New York:
% TELOS/Springer--Verlag.

% [3] Hasselmo, M.E., Schnell, E.\ \& Barkai, E.\ (1995) Dynamics of learning and
% recall at excitatory recurrent synapses and cholinergic modulation in rat
% hippocampal region CA3. {\it Journal of Neuroscience} {\bf 15}(7):5249-5262.
% }

%%%%%%%%%%%%%%%%%%%%%%%%%%%%%%%%%%%%%%%%%%%%%%%%%%%%%%%%%%%%
\appendix

\appendix

\section{Appendix}

\subsection{Limitations and Future Work}
\label{sec:appendix_limitations}

While \method{} delivers substantial token and latency savings, we identify four core systemic constraints:

\begin{itemize}
    \setlength{\itemsep}{1pt}
    \item \textbf{Inference Slack Availability:} \method{} targets multi-turn loops with high tool usage. On conversational workloads with minimal tool calls, the available slack approaches zero. Premature truncation in these environments can occasionally induce an extra trailing turn, degrading performance.
    \item \textbf{Upstream Dispatch Inefficiencies:} Our framework optimizes inference only after an action is actively launched. It cannot resolve structural errors where the initial speculative fork triggers an unnecessary tool call that the main branch ultimately rejects (accounting for 14.9\% of dispatches on $\tau^2$-bench).
    \item \textbf{Complex Argument Synthesis:} The generation of complex, free-form, or nested tool arguments delays sequence stabilization ($A_j$). This extended token distance narrows the available optimization window between prediction consensus and tool readiness ($t_{\text{ready}}$).
    \item \textbf{vLLM Scheduler Interference:} Executing intermediate auxiliary token probes alters the runtime batch composition inside vLLM. While prefix-cache hits keep processing overhead bounded, high concurrent traffic introduces minor scheduling friction that requires deep algorithm-system co-design.
\end{itemize}

\subsection{Glossary}
\label{glossary}
\medskip
\noindent
\begin{center}\footnotesize
\begin{tabular}{@{}ll@{\hskip 16pt}ll@{}}
\toprule
$H$ & turns per episode & $c,\theta$ & span confidence, dispatch threshold \\
$T$ & wall-clock  & $p,\tau$ & free-probe signal, truncation threshold \\
$\bar{d}$ & mean per-token decode latency & $\alpha$ & speculation acceptance rate \\
$T^{\mathrm{tool}}_t$ & tool execution time at turn $t$ & $\sigma_t$ & per-turn slack (\S\ref{sec:method}) \\
$f_{\mathrm{tool}}$ & tool time as a fraction of $T$ & $f_{\mathrm{slack}}$ & slack as a fraction of $T$ \\
\bottomrule
\end{tabular}
\end{center}

\subsection{Worked Example}
\label{appendix:example}

This four-turn task illustrates successful truncation and two cases in which a
probe-side guard prevents an incorrect exit. The user asks to exchange a bookshelf for
a camera. The agent must authenticate the user, retrieve the order, determine that it
has not yet been delivered, and decline the exchange. The final column reports either
the token at which generation was truncated or the guard that prevented truncation.

\begin{center}\footnotesize
\setlength{\tabcolsep}{4pt}
\begin{tabular}{@{}clllc@{}}
\toprule
turn & main stream emits & forced probe predicts & $p$ over probes & outcome \\
\midrule
1 & \texttt{find\_user\_id}
  & \emph{same}, $\times 3$
  & $.999\;\;1.00\;\;1.00$
  & \textbf{cut} at token 173 \\
2 & \texttt{get\_user\_details}
  & \emph{same}, $\times 3$
  & $.979\;\;1.00\;\;1.00$
  & \textbf{cut} at token 158 \\
3 & \texttt{get\_order(\#\ldots037)}
  & \texttt{\#\ldots222}, then \texttt{\#\ldots037}
  & $.986 \ldots 1.00$
  & dispatch mismatch \\
4 & \emph{message, no call}
  & \texttt{transfer\_to\_human}, $\times 5$
  & $.000 \times 5$
  & $p < \tau$ \\
\bottomrule
\end{tabular}
\end{center}

Turns~1 and~2 show the intended behavior. In both cases, the prediction is exact and
quickly becomes stable. 
The exit therefore arms at probe~2 and fires at the next probe boundary. 
%This occurs $4.1$\,s after the required observation becomes available, so the
% deadline is not the limiting condition. 
Truncation reduces chain-of-thought length from
$386$ to $176$ tokens on turn~1 and from $493$ to $160$ tokens on turn~2. Together,
the two turns take $20.8$\,s, compared with $46.9$\,s without truncation, including
continuation costs of $2.1$\,s and $1.5$\,s.

Turns~3 and~4 show why the remaining guards are necessary. On turn~3, probe~0 predicts
the nonexistent order ID \texttt{\#W5362222}, and a request (tool-call) for that order is dispatched.
Probes~1-3 then converge on the correct ID, \texttt{\#W5362037}, with $p = 1.000$.
Although the stability and free-probe conditions are satisfied, the dispatch-match guard
blocks the exit because the in-flight request concerns a different, fictitious order.
Truncating at this point would therefore provide no useful result.

On turn~4, the forced probe repeatedly predicts
\texttt{transfer\_to\_human\_agents}, producing the same 54-word summary five times.
From probe~2 onward, this prediction is both stable and identical to the dispatched
call, so stability and dispatch matching alone would permit an exit. The main stream,
however, ultimately sends a message to the user without making any tool call. The free
probe correctly blocks truncation: it assigns $p = 0.000$ at every probe, with
\texttt{'YOU'} as the top token at probability $0.999$. This example makes the gap
between the forced probe's conditional prediction and the main stream's marginal
decision concrete.

\subsection{Probe Cadence}
\label{app:cadence}

Probe~0 is issued when the main stream emits its first token. Each subsequent probe is
issued after the stream decodes another $\Delta = 50$ CoT tokens and reaches
a sentence boundary, identified by a trailing \texttt{\textbackslash n},
\texttt{".\,"}, or \texttt{".\textbackslash n"}. If no boundary occurs within an
additional $\Delta_g = 30$ tokens, the probe is issued immediately. We issue at most
$J = 15$ probes per turn. Thus, the cadence is primarily token-driven; sentence
boundaries determine where near the target interval a probe is placed. Each probe
requires two prefix-cache-hit requests: a forced probe of at most $160$ tokens and a
one-token free probe.

Previous works stop probing after its first confident dispatch.
% This behavior provides only one post-dispatch sample, making it impossible to evaluate
% stability over $k \ge 2$ samples. 
We therefore continue probing after dispatch. Exit mechanism disabled (ours (off) in \S\ref{sec:results}) incurs the full cost of these additional probes while
disabling truncation,
% allowing the treatment comparison to 
which isolate the benefit of the
exit rule.

% \subsection{Probe cadence}
% \label{app:cadence}
% % \paragraph{Probe cadence.}
% Probe $0$ fires at the main stream's first streamed token. Probe $j>0$ fires once the
% main stream has decoded a further $\Delta = 50$ chain-of-thought tokens \emph{and} the
% text sits at a sentence boundary (a trailing \texttt{\textbackslash n},
% \texttt{".\,"} or \texttt{".\textbackslash n"}); if no boundary arrives within
% $\Delta_g = 30$ additional tokens it fires anyway. At most $J = 15$ probes are issued
% per turn. The cadence is token-driven rather than sentence-driven, since most sentences
% are shorter than $\Delta$. Each probe is two requests --- the forced probe
% ($\le 160$ tokens) and the one-token free probe --- both prefix-cache hits.

% This departs from the underlying dispatch controller in one respect that the rule
% forces rather than chooses: that controller stops probing at its first confident
% dispatch, so a stability criterion over $k \ge 2$ samples could never be evaluated.
% We continue probing after dispatch. The resulting cost is charged in full to the
% mechanism by our null control (\S\ref{sec:experiments}), which pays it and truncates
% nothing.

\begin{figure}[t]
    \centering
    \includegraphics[width=\linewidth]{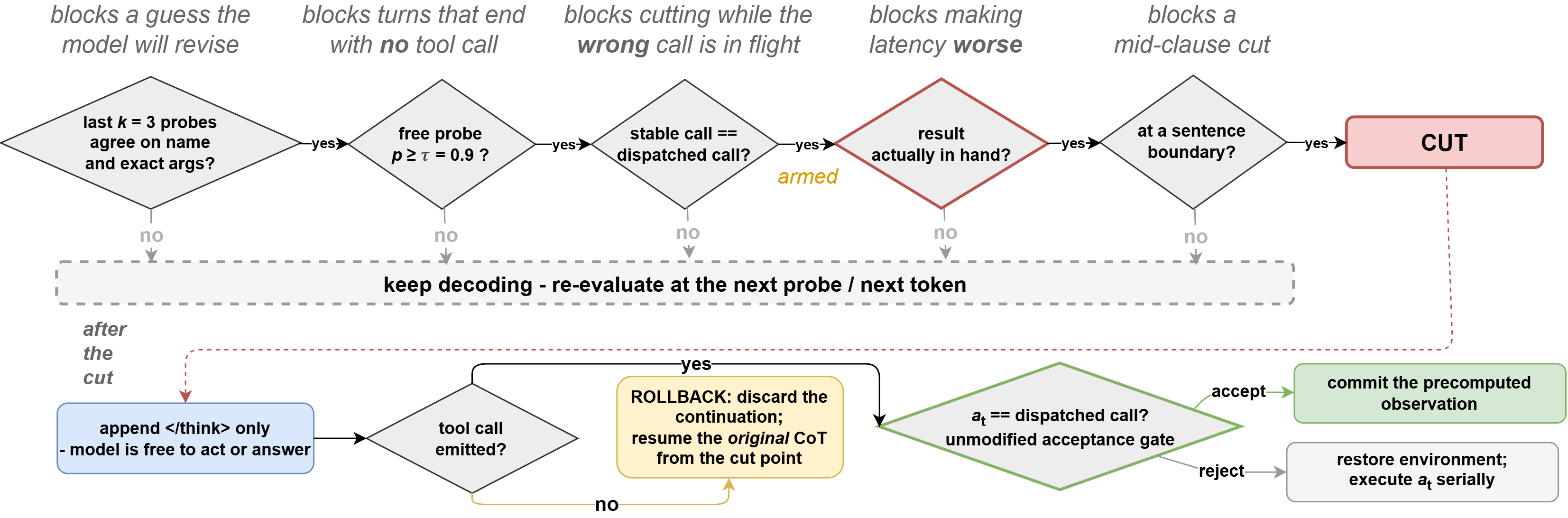}
    \caption{Overview of the decision mechanism for terminating main-stream reasoning and the failure mode blocked by each guard.
    The probe prediction is never adopted.
    After truncation, the model derives its own action, which must pass the unmodified acceptance gate of main stream.
    A turn that declines to make a tool-call is rolled back onto its original reasoning.}
    % Measured timeline of a real
    % truncated turn. A probe forked from the main stream predicts the upcoming action
    % and dispatches it early; the observation is ready at $t^{\mathrm{ready}}$, but the
    % main stream would otherwise keep reasoning for another $12$\,s. Once three
    % consecutive probes agree, the free probe reads $p \ge \tau$, and the stable
    % prediction still equals the dispatched call, the exit \emph{arms}; it \emph{fires}
    % at the next sentence boundary, which cannot precede $t^{\mathrm{ready}}$ by
    % construction. \emph{(b)} The decision, with the failure each guard blocks. The
    % probe's prediction is never adopted --- after the cut the model re-derives its own
    % action, which still passes the unmodified acceptance gate, and a turn that declines
    % to act is rolled back onto its original reasoning.}
    \label{fig:overview_mechanism}
\end{figure}

\subsection{Design of the Exit Mechanism}
\label{app:exit_points}

Figure~\ref{fig:overview_mechanism} summarizes the exit mechanism. Three design choices in
Eqs.~\eqref{eq:fire} and~\eqref{eq:continuation} are central to its behavior.

\emph{First, readiness is observed rather than estimated.}
The controller evaluates $\mathrm{ready}_t$ by polling whether the speculative tool
execution has completed. The deadline is therefore exact and requires no prior model of
tool latency. In the example in Figure~\ref{fig:overview}(a), the tool result is
available well before the exit rule permits truncation. At this latency, the mechanism is
limited by prediction stability rather than tool readiness; the deadline begins to bind
only as $T^{\mathrm{tool}}$ increases.

\emph{Second, the continuation remains unconstrained.}
After truncation, we append only \thinkend{} and allow the model to generate its own
action; we never inject the dispatched call $\check a_t$. The mechanism therefore cannot
manufacture an action on the model's behalf, and action-level quality continues to
measure the model's behavior.

\emph{Third, the exit guards do not replace the acceptance gate.}
Equation~\eqref{eq:arm} compares the probe prediction with the dispatched call. The
acceptance gate separately compares the model's final call with the dispatched call.
Thus, truncation cannot force acceptance: if the continuation emits a different action,
the speculative result is discarded and the emitted action is executed normally.

\paragraph{No-tool rollback.}
A continuation may emit no tool call even after satisfying all exit guards, indicating
that the free probe incorrectly predicted a tool-emitting turn. In this case, truncation
may have interrupted a terminal answer. We found no reliable way to identify such cases
at exit time: across $104$ truncations, harmful exits were not separable by token count,
$p$, confidence, or probe index, and candidate thresholds rejected substantially more
beneficial exits than harmful ones.

We therefore detect this outcome directly and roll it back without a threshold. The
controller discards the continuation and resumes the original reasoning from
$n^{\dagger}$ without appending \thinkend{}. Because this is a warm-cache continuation,
it only regenerates tokens that the untruncated stream would otherwise have produced.
The mechanism can consequently incur additional latency, but a no-tool continuation is
not committed as the turn's final output.

\paragraph{Operating point.}
We selected $k = 3$, $\tau = 0.9$, and exact dispatch matching offline, before the
full task scale evaluations. We replayed recorded probe streams under candidate rules and
measured two quantities: \emph{precision}, the fraction of proposed truncations whose
predicted call matched the call produced by the untruncated turn; and \emph{harvest}, the
wall-clock time those truncations would save. On $\tau^2$-bench, the selected operating
point achieved $0.917$ precision while targeting $34.4\%$ of wall-clock time. We then
froze the rule and applied it without retuning to HotpotQA and BFCL v4 Multi-Turn, where
precision increased to $0.938$ and $0.976$, respectively.

\begin{algorithm}[t]
\caption{One agent turn under \method}
\label{alg:turn}
\begin{algorithmic}[1]
\Require context $C_t$; thresholds $\theta$ (dispatch), $\tau$ (truncation); stability $k$;
         cadence $\Delta,\Delta_g$; probe budget $J$
\State launch main stream $\pi(\cdot \mid C_t)$ with a cooperative stop hook
\State $\check a \gets \varnothing$; \; $\mathcal{P} \gets [\,]$ \Comment{dispatched call; probe history}
\State \textbf{wait} for first streamed token
\For{$j = 0,\dots,J-1$}
    \If{$j > 0$} \textbf{wait} until $\Delta$ further CoT tokens \textbf{and} sentence boundary (or $\Delta+\Delta_g$ tokens) \EndIf
    \State $r_{1:n_j} \gets$ snapshot of main's CoT
    \State $(\hat a^{(j)}, c^{(j)}) \gets \Call{ForcedProbe}{C_t, r_{1:n_j}}$ \Comment{cache hit}
    \State $p^{(j)} \gets \Call{FreeProbe}{C_t, r_{1:n_j}}$ \Comment{1 token, logprobs=20, Eq.~\eqref{eq:ptool}}
    \State append $(\hat a^{(j)}, c^{(j)}, p^{(j)})$ to $\mathcal{P}$
    \If{$\check a = \varnothing$ \textbf{and} $c^{(j)} \ge \theta$}
        \State $\check a \gets \hat a^{(j)}$; \; \textsc{Snapshot}(env); \; $F \gets \Call{DispatchAsync}{\check a}$
    \EndIf
    \If{$A_j$ holds (Eq.~\eqref{eq:arm})} \State $\mathrm{armed} \gets \textsc{True}$; \textbf{break} \EndIf
\EndFor
\State \textbf{on each streamed token $n$:} \textbf{if} $\mathrm{armed} \wedge F.\textsc{done}() \wedge \mathrm{bnd}(r_{1:n})$ \textbf{then} stop main at $n^{\dagger} \gets n$
\If{main was truncated}
    \State $a_t \gets$ \Call{Complete}{$C_t, r_{1:n^{\dagger}} \oplus \thinkend$} \Comment{free continuation}
    \If{$a_t = \varnothing$} \State $a_t \gets$ \Call{Complete}{$C_t, r_{1:n^{\dagger}}$} \Comment{rollback: resume original CoT} \EndIf
\Else \State $a_t \gets$ main's own emitted action
\EndIf
\State \textbf{if} $a_t \equiv \check a$ \textbf{then} commit $F$'s observation \textbf{else} \textsc{Restore}(env); execute $a_t$ serially
\end{algorithmic}
\end{algorithm}

\paragraph{Speculation with state-mutating tools.}
Prior speculative-dispatch methods are limited to read-only tools because an incorrect
speculative call could alter the state on which the run is evaluated. Two of our
workloads instead use mutable backends, including a file system and a trading account.
Before dispatch, we snapshot the backend and execute the speculative call on an isolated
copy. The resulting state is committed only if the acceptance gate matches the
speculative call to the model's final call; otherwise, the original state is restored
(Alg.~\ref{alg:turn}, lines 11 and 20). Speculation therefore affects only latency until
acceptance, extending the mechanism to state-mutating tools without relaxing the
acceptance criterion.

\subsection{Non-inferiority analysis}
\label{app:noninf}

We evaluate action match using a paired, task-level percentile bootstrap with
$20\,000$ resamples. Resampling tasks preserves within-task dependence; treating turns
as independent would produce intervals that are too narrow. For each resample, we
compute
\[
\frac{\sum_i m_i^B-\sum_i m_i^A}{\sum_i g_i},
\]
where $m_i^A$ and $m_i^B$ are matched actions and $g_i$ is the number of golden
actions for task $i$. Resampling both numerator and denominator avoids giving equal
weight to tasks with different numbers of actions.

We use action match rather than task success because $\tau^2$ assigns the same
task-level score to trajectories matching 9 of 10 actions and 0 of 10. Across the five
arms, only $26$--$30$ of $155$ tasks succeed, whereas action match provides $692$
scored events and approximately $4.5\times$ greater resolution.

\paragraph{Calibrating the margin.}
We estimate the apparatus noise from comparisons that should not change model decisions.
Before examining the treatment effect, we set the non-inferiority margin to the largest
absolute difference among these calibrators:

\begin{table}[h]
\centering\footnotesize
\begin{tabular}{@{}lrrl@{}}
\toprule
calibrator & $\Delta$ actions & $\Delta$ pp & expected equivalence \\
\midrule
serial $\to$ SPORK
  & $+22$ & $\mathbf{+3.18}$ & lossless speculation \\
serial $\to$ SPORK(retry)
  & $+14$ & $+2.02$ & unchanged acceptance gate \\
SPORK $\to$ SPORK(retry)
  & $-8$ & $-1.16$ & unchanged acceptance gate \\
SPORK(retry) $\to$ ours (off)
  & $-17$ & $-2.46$ & instrumentation only \\
\bottomrule
\end{tabular}
\end{table}

These differences arise from serving nondeterminism, including changes in batching,
KV-cache state, and floating-point reduction order. We therefore set
$\delta=3.18$ pp. Effects smaller than this margin cannot be
resolved reliably by the present experiment.

\paragraph{Non-inferiority results.}
A comparison is non-inferior when its one-sided $95\%$ lower confidence bound (LCB)
exceeds $-\delta$.

\begin{table}[h]
\centering\footnotesize
\begin{tabular}{@{}lrrl@{}}
\toprule
comparison & $\Delta$ pp & 95\% LCB & verdict \\
\midrule
\textbf{SPORK(retry) $\to$ ours (on)}
  & $+2.02$ & $-2.22$ & \textbf{non-inferior} \\
serial $\to$ ours (on)
  & $+4.05$ & $+0.78$ & non-inferior; LCB $>0$ \\
SPORK $\to$ ours (on)
  & $+0.87$ & $-3.11$ & non-inferior \\
ours (off) $\to$ ours (on)
  & $+4.48$ & $+0.83$ & non-inferior; LCB $>0$ \\
\bottomrule
\end{tabular}
\end{table}

All treatment comparisons satisfy non-inferiority at $\delta=3.18$\,pp. Against
SPORK(retry), our method is $1.33\times$ faster with an action-match difference of
$+2.02$\,pp and an LCB of $-2.22$\,pp. We therefore conclude that any quality loss is
smaller than the experiment can resolve.
\subsection{Per-task distributions and mechanism behavior}
\label{app:mech}

\paragraph{Per-task speedup.}
Against SPORK(retry), our baseline, the median task-level speedup is
$\mathbf{1.28\times}$ (P05 $0.43\times$, P25 $1.01\times$, P75 $1.91\times$,
P95 $4.40\times$). Our method is faster on $118/155$ tasks ($76\%$), including
$34$ tasks with at least a $2\times$ speedup. It is slower on $37$ tasks ($24\%$),
including $9$ by at least $2\times$. The worst case is \texttt{airline\_7} at
$0.21\times$; the largest absolute regression is \texttt{retail\_113} at $+305$\,s.

\begin{table}[h]
\centering\footnotesize
\begin{tabular}{@{}lrrrrrrr@{}}
\toprule
arm & P10 & P25 & P50 & P75 & P90 & P95 & P99 \\
\midrule
serial       & 80 & 109 & 163 & 233 & 421 & 510 & 667 \\
SPORK        & 91 & 121 & 174 & 250 & 412 & 478 & 542 \\
SPORK(retry) & 90 & 115 & 170 & 237 & 393 & 457 & 540 \\
\textbf{ours (on)}
             & \textbf{64} & \textbf{90} & \textbf{125} & \textbf{197}
             & \textbf{267} & \textbf{412} & \textbf{484} \\
\bottomrule
\end{tabular}
\end{table}

\begin{figure}[h]
    \centering
    \includegraphics[width=\linewidth]{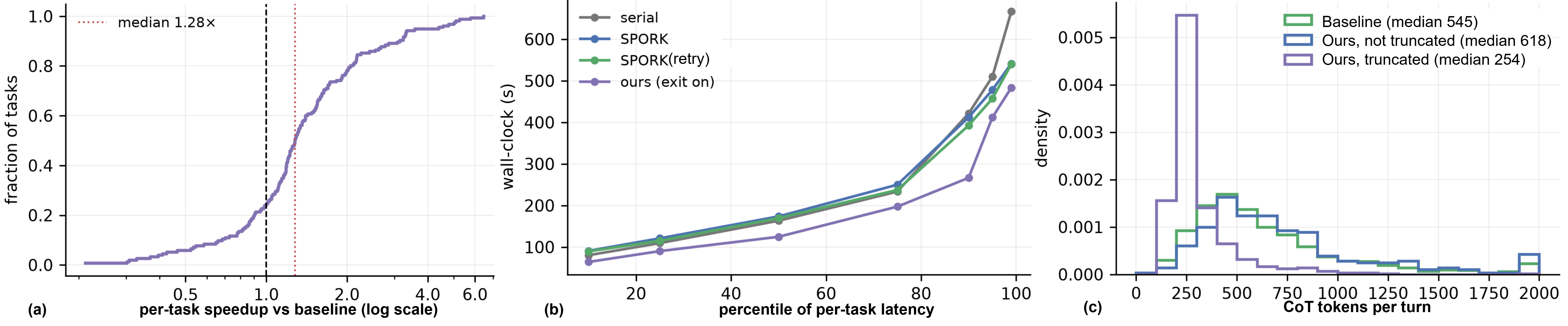}
    \caption{\textbf{Task-level latency and truncation behavior.}
    \emph{(a)} Distribution of per-task speedup over SPORK(retry), with a median of
    $1.28\times$. \emph{(b)} Task-latency percentiles by arm; our method reduces
    latency throughout the distribution. \emph{(c)} Chain-of-thought length for
    SPORK(retry) and for truncated and non-truncated turns. Truncation targets turns
    on which the action stabilizes early.}
    \label{fig:result-appendix}
\end{figure}

The improvement extends across the latency distribution. As
Figure~\ref{fig:result-appendix}(a--b) shows, relative to SPORK(retry), P90 falls from
$393$ to $267$\,s ($-32\%$), and P95 falls from $457$ to $412$\,s. This is an
aggregate result, not a per-task guarantee: the deadline applies per turn, and
truncation can change the trajectory or add turns.
% (\S\ref{sec:results:costs}).

\paragraph{Where truncation occurs.}
Figure~\ref{fig:result-appendix}(c) shows that truncated turns are substantially
shorter than the SPORK(retry) baseline:

\begin{table}[h]
\centering\footnotesize
\begin{tabular}{@{}lrrrr@{}}
\toprule
group & $n$ & mean & P50 & P90 \\
\midrule
SPORK(retry), all turns & 839 & 708 & 545 & 1\,150 \\
ours, not truncated     & 282 & 854 & 618 & 1\,340 \\
ours, truncated         & 589 & 307 & 254 & 490 \\
\bottomrule
\end{tabular}
\end{table}

Across all turns, our method emits $484$ chain-of-thought tokens per turn, compared
with $708$ for SPORK(retry) ($-32\%$). Despite using $3.8\%$ more turns, it reduces
total chain-of-thought tokens by $29\%$ and total decoded tokens by $27\%$.
Non-truncated turns are longer than the baseline average, suggesting that the exit
selectively fires when the model settles on an action early.

\paragraph{Dispatch outcomes.}
The following outcomes partition all dispatched speculations:

\begin{table}[h]
\centering\footnotesize
\begin{tabular}{@{}lrrr@{}}
\toprule
outcome & SPORK & SPORK(retry) & ours (on) \\
\midrule
accepted ($\alpha$): name and arguments match
  & 492 (61.3\%) & 531 (65.2\%) & \textbf{567 (68.6\%)} \\
rejected: main stream emits a different call
  & 173 (21.6\%) & 147 (18.1\%) & 137 (16.6\%) \\
rejected: main stream emits no call
  & 137 (17.1\%) & 136 (16.7\%) & 123 (14.9\%) \\
total
  & 802 & 814 & 827 \\
\bottomrule
\end{tabular}
\end{table}

The exit guards directly address mismatched calls. No-call rejections remain a separate
limitation: the speculative path dispatches a tool, but the main stream ultimately
responds to the user without calling one.

\subsection{Offline selection of the exit rule}
\label{app:oracle}

We selected the exit rule by replaying recorded probe streams under candidate settings.
Each setting was evaluated on \textbf{precision}, the fraction of proposed truncations
whose predicted call matched the untruncated call, and \textbf{harvest}, the fraction of
wall-clock time it could remove. The selected rule ($k=3$, $\tau=0.9$, with exact
dispatch matching) achieved $0.917$ precision and targeted $34.4\%$ of wall-clock time
on $\tau^2$-bench. We fixed this rule before the full evaluations and performed no
online tuning.

\paragraph{Transfer without retuning.}
The same rule achieved precision of $0.938$ on HotpotQA and $0.976$ on BFCL. Precision
increased as arguments became easier to predict: $\tau^2$ primarily uses identifiers
already in context, HotpotQA composes free-text queries, and BFCL uses enums, numeric
values, and identifiers from prior observations.

\newpage

\end{document}